# RNA control of HIV-1 particle size polydispersity

Authors: Cendrine Faivre-Moskalenko[1], Julien Bernaud[1], Audrey Thomas[2,3], Kevin Tartour[3], Yvonne Beck[1], Maksym Iazykov[1], John Danial[1], Morgane Lourdin[3], Delphine Muriaux[2,3]* & Martin Castelnovo[1]*.

*Co-last authors

1 Université de Lyon, Laboratoire de Physique, Ecole Normale Supérieure de Lyon, CNRS, Lyon, FRANCE
2 CNRS UMR 5236, Centre d'études d'agents Pathogènes et Biotechnologies pour la Santé , Montpellier , FRANCE.
3 Université de Lyon, INSERM, U758, Ecole Normale Supérieure de Lyon, Lyon, FRANCE

**Abstract:**

HIV-1, an enveloped RNA virus, produces viral particles that are known to be much more heterogeneous in size than is typical of non-enveloped viruses. We present here a novel strategy to study HIV-1 Viral Like Particles (VLP) assembly by measuring the size distribution of these purified VLPs and subsequent viral cores thanks to Atomic Force Microscopy imaging and statistical analysis. This strategy allowed us to identify whether the presence of viral RNA acts as a modulator for VLPs and cores size heterogeneity in a large population of particles. These results are analyzed in the light of a recently proposed statistical physics model for the self-assembly process. In particular, our results reveal that the modulation of size distribution by the presence of viral RNA is qualitatively reproduced, suggesting therefore an entropic origin for the modulation of RNA uptake by the nascent VLP.

**Introduction**

Viral life cycle has been investigated for several decades, and significant progresses in the understanding of its different steps at the molecular level have been achieved[1,2]. Different viruses from different families show in particular distinct self-assembly pathways. In the case of non-enveloped viruses, the protein self-assembly is performed in the cytosol of the host cell. *In vitro* reconstitution by mixing recombinant capsid proteins with or without their nucleic acid under various environmental conditions allows one to show that capsid sizes and morphologies are not unique[2,3]. However, the self-assembly conditions met in the host cell select a particular capsid morphology, and give rise to relatively small polymorphism and size heterogeneity *in vivo*[4].

In contrast to this observation, retroviruses exhibit quite a large polymorphism and size polydispersity *in vivo* for similar growth conditions[5]. The particular complex assembly pathway for retroviruses is likely to be responsible for such a polydispersity, although this has never been firmly established. The HIV-1 assembly scenario is composed of at least three consecutive steps: the multimerization of the viral Gag proteins at the plasma membrane, the budding of the viral particle and its subsequent maturation. During the budding step, the viral Gag proteins, which are the major constituent of the viral particle, self-assemble at the plasma membrane in a cooperative way with the viral dimeric RNA[6]. Interestingly, some cellular RNA's of the host cell have been shown to be incorporated as well within the nascent particle[7,8]. As with other RNA viruses, the HIV-1 genomic RNA bears a specific Packaging Signal (PSI or "ψ") enhancing its preferential uptake into VLPs[9,10]. The assembly site will buckle out of the bilayer host cell membrane as more proteins are incorporated, leading to an almost spherical particle connected to the bulk membrane through a thin neck[11]. The recruitment of additional protein complexes from the ESCRT family allows the liberation of the particle[12]. During the maturation step, the HIV-1 Gag protein is processed by the viral protease into its component domains MA, CA and NC, and the linker peptides sp1, sp2 and p6, resulting in significant structural changes of the virion[13].

Both the budding and maturation steps have been studied independently with various assays[13,14]. These *in vitro* studies using specific or non specific nucleic acids have provided important clues about the retroviral assembly pathways. However, the global understanding is far from being complete. In particular there are at least two unanswered questions about self-assembly mechanism: *(i)* what is the origin of the large size polydispersity observed for viral particles produced by infected cells, and *(ii)* what are the molecular factors that control this phenomenon? In order to address these questions, the morphological properties of purified virus-like particles (VLP) and mature viral cores produced by cells were investigated.

In the present work, we propose and illustrate a novel strategy in order to make progress in the identification of the main determinants of virion morphology and size variability. The idea is to be able to quantify precisely the size distribution of viral particles produced by cells, and to use this size heterogeneity as a global reporter for HIV-1 assembly within cells. We applied this strategy by using a unique combination of retrovirus purification protocols, biochemical tools, Atomic Force Microscopy (AFM) imaging at the single virus level and automated image analysis. AFM was used in this study for its capacity to image a large number of individual particles: the adsorption of the molecular objects on a surface prior to AFM imaging increases significantly the number of objects effectively probed. Moreover the automated analysis of topographic images is facilitated as compared to other technique like cryo-EM[15]. AFM also allows to probe the mechanical properties of the particles, as it was recently done on HIV-1 derived particles by Kol and coworkers[16],[17]. Using AFM in order to image viruses, we analyzed the statistical properties of large populations of HIV-1 derived virus-like particles (VLP) and cores in the presence or in the absence of viral RNA. Using this approach, we showed that the presence of viral RNA is one of the determinants of the VLP size polydispersity: particles that contain viral RNA are smaller on average and more homogeneous in size in comparison with particles that packaged other cellular RNAs in place of the viral genome. These observations are interpreted using a general modeling of the viral particle self-assembly process. The model is based on the classical description used to address quantitatively the thermodynamical properties of micellization[18]. Within this approach, it is found that the RNA control of HIV-1 size polydispersity observed in the experiments is qualitatively reproduced. This strongly suggests that the observed control of particle size is associated to entropic effects during the self-assembly.

## Results

*Biochemical characterization*
The first step prior to size characterization was to produce and purify virus-like particles (VLP) and viral cores according to the procedure described in the *Material and Method* section. In order to check whether the combined steps of particle production and purification led to biologically relevant VLPs and cores, we first analyzed their protein content by performing immunoblot analysis (Fig.1a).

For mature VLPs, pr55 Gag cleavage products (CAp24, MAp17) were mainly detected by immunoblots with anti-MA and anti-CA antibodies (Fig. 1a, lane 1 "VLP"). The viral enzymes integrase (IN) and reverse transcriptase (RT) were also found in mature VLPs as shown by immunoblots with the corresponding antibodies (Fig. 1a, lane 1"VLP", IN and RT). In contrast, in purified cores, while the CA proteins were present, the MA protein as part of the viral membrane disappeared in the isolated cores and no immature p55Gag was detected (Fig. 1a, lane 2 "cores") as expected since MA remains attached to the viral membrane. RT and IN were also found in the mature cores. The two subunits of RT (p66 and p51) and Integrase were detected in mature VLPs and cores, with a lower intensity for cores, indicating that during triton purification of cores from VLPs some instable cores or particles get lost. Immunoblots of VLPs and cores prepared without viral ψRNA were similar to those shown in figure 1 a. In addition, to further check that VLP and cores were functional after purification, their reverse transcriptase activities were estimated (Fig.1b). As expected, RT activity was detected in both VLP and cores (+ or – viral ψRNA), with about half or one third less efficacy for the purified mature cores, in agreement with immunoblot analysis.

*AFM imaging and particle characterization*
The purified samples were then deposited on poly-L-lysine-coated mica surfaces and dried for air-imaging with Atomic Force Microscope (see *Material and Methods*). Simple visual analysis of VLPs and cores by AFM images confirms the presence of objects with appropriate sizes and morphological features. This simple consistency check was performed for each purification. Examples of typical mature VLP and core images are shown respectively in figure 1c and 1d. One can recognize spherical shape VLP of about 100nm in average diameter (Fig 1c) and typical HIV conical shape cores (Fig 1d) with an average length of 75nm and width of 50nm. In order to be able to sort in an objective way different populations within one single purification condition based on morphological features, we developed home-made automated image analysis scripts (see *Material and Methods*). These tools enable the sorting of objects according to the simultaneous consideration of their height and lateral size (Figure 2). VLPs are automatically detected on images by focusing on particles that have both a minimal height (typically 25nm) and a minimal lateral size (typically 90nm in diameter). In the case of purification containing cores, these thresholds have been adapted in order to focus on this population.

The statistical comparison of size and shape parameters for VLP and cores is shown in figure 3. The clear difference in asymmetry of VLP and cores observed on the selected particles of figure 1b and 1c is quantified at a statistical level using 2D histograms of long diameter as function of short diameter for each particle (see also Figure S3 in supplementary material). It is observed that the size distribution of VLP has a large dispersion, and that this dispersion is isotropic, *i.e.* the distribution is smeared along the line for which the short and the long diameter are equal. The dispersion (between 10 and 20 nm) is relevant to the self-assembly process, since it is larger than the resolution associated to the convolution of samples with tip size (roughly 5 nm). The comparison of this large dispersion with dispersion measured for other non-enveloped RNA viruses (typically between 5 and 10 nm) indicates indirectly that size control during the self-assembly process is not as stringent for HIV-1[4,19]. On the other hand, the dispersion of particles from core purification is smaller and it is essentially observed on the long axis of the particles. Indeed, the large dispersion observed for VLP is partially transmitted to the long axis of the conical capsid (between 10 and 20 nm). The dispersion along the minor axis is strongly reduced, reflecting a more stringent control of this particular dimension of the conical capsid (between 5 and 10 nm). Our measurements are in quantitative agreement with previous cryo-EM characterization by Briggs *et al*[20].

### *The presence of viral RNA modulates VLP size polydispersity*

Having established the size distribution of HIV-1 VLPs, we compare these distributions for particles differing in viral RNA content. These particles are produced by two distinct methods: in the first method, VLP are produced using a single transfection of a plasmid encoding Gag/Gag Pol proteins in a lentiviral vector[21]. In the second method, the particles are produced by adding in the same transfection step another plasmid expressing a non infectious HIV-1 ψRNA[22]. As a consequence, the second type of particles contains mainly viral ψRNA (" VLP +ψ ") , while the first type mainly contain cellular RNAs in place of the viral RNA (" VLP -ψ "), as it has been previously described [7,8]. Interestingly, it has been previously shown that VLP +ψ particles do contain also a significant amount of cellular RNAs. We first checked that the size distribution of VLP +ψ is not significantly dependent on the amount of plasmid expressing viral RNA used for the transfection (Figure S4 in the in supplementary material.). Next, the comparison of the size distribution for the two types of particles +ψ and -ψ are shown in figure 4. In the case of VLP, both the peak position of the 2D size distribution (long diameter/short diameter) and its dispersion are changed: VLP +ψ particles are smaller on average and less polydisperse (average diameter= 98nm and standard deviation=7nm) than VLP -ψ particles (average diameter= 106nm and standard deviation=11nm). A shift in the 2D size distribution is also observed for cores purified from the two previous types of particles, although with a lower amplitude. These results are summarized in figure 4e by using a box plot representation. Both in the case of VLP and cores, a statistical test confirm that the size distributions from -ψ and +ψ are different (p-values=$2.10^{-7}$ for VLP and $7.10^{-10}$ for cores) (Fig.4e). In addition, since we did not find significant differences between size distributions of +ψ for different plasmid expression in our experimental conditions (figure S4 in supplementary material), this suggests that the change in particle size distribution is rather an all-or-none effect requiring only a small amount of viral +ψ plasmid expression. Overall, our results suggest that from a statistical point of view the presence of a viral RNA allows the significant modulation of HIV-1 particle size variability. Interestingly, we also checked that AFM imaging in liquid environment, although much difficult to achieve, gives essentially the same results about the role of viral RNA on VLP morphology. Indeed, the relative effect of size average and polydispersity of VLP population upon the presence of viral RNA is consistent. These results are available in the supplemental material (figure S5 in supplementary material).

### *Thermodynamics of viral assembly in the presence of viral and cellular RNAs*

It is possible to understand with very general physical arguments how RNA might influence the size distribution of viral particles. More precisely, we present below a simple statistical model of virus self-assembly that is able to reproduce the main observation of our study: the presence of viral RNA is significantly favoring smaller size particles. The derivation of the model is postponed to the supplemental data, and qualitative physical arguments summarizing the results of the model are proposed below. Moreover, the reader is referred to a recently published thorough analysis of the self-assembly equations for more details[23].

First, it must be realized that the process of viral proteins self-assembly (in the absence of viral RNA) in order to form capsids is controlled by the balance between enthalpic and entropic effects: on the one hand the enthalpy represents the free energy gained by individual proteins at forming each capsid, and on the other hand the entropy represents the number of ways to assemble these capsids at constant internal energy[18,19,24] . At this step, an interesting observation can be made : if the enthalpy is only weakly selective in capsid size, the entropy will favor smaller particles, because more « small » particles can be made than « large » particles at constant number of total proteins, leading to larger entropy for the former and smaller entropy for the latter. Now we claim that the uptake of RNA inside viral particles allows to modulating the population ratio of these particles thanks to similar entropic effects. More precisely, the consequence of the presence of multiple types of RNA during the self-assembly of capsids is two fold: first, entropy will favor the uptake of longer RNAs like viral genomes, and second, their capsid preferred size is shifted to smaller values. These behaviors are illustrated below.

As we already mentioned it, the two types of particles of interest in our study are VLP +$\psi$ particles, which are thought to contains mainly viral RNA, and also some cellular RNAs, and VLP -$\psi$ which are thought to contain mainly cellular RNAs[8] . In the study by Rulli and co-workers, the amount of RNAs in each particle has been quantified, and it turns out that the total amount of RNA is roughly the same in these two types of particles. Such a linear relation between the total nucleotides of RNA and the number of proteins in the capsid is generally observed for RNA viruses[18,25]. Recent works suggest that this feature is based on the electrostatic interactions between proteins in the capsid and RNA. As a consequence, in VLP -$\psi$ particles, some cellular RNAs present in the cytoplasm replace the viral RNA in order to maintain a constant level of RNA in each particle. Since viral RNA is known to have a larger nucleotide length than most of cellular RNAs this has a strong impact on the number of ways capsid can be assembled in their presence, or equivalently on the total configuration entropy for capsid formation. Indeed, the replacement of several « small » cellular RNAs by the two « large » dimerized viral RNAs allows the formation of a larger number of VLP +$\psi$ particles at constant amount of cellular RNAs, as the effective number of cellular RNAs is reduced in VLP +$\psi$ particles. As a consequence, VLPs containing *large* viral RNA are intrinsically more favorable in terms of the entropic cost for the self-assembly than VLPs containing only cellular RNAs.

This is illustrated by numerical calculations based on the simple micellization thermodynamic model which has been used in order to describe the aggregation properties of surfactant solutions and which is presented in the Material and methods section. The key idea of the calculation is to restrict the product of the self-assembly to two final states: two particles of the same size, but of different RNA content (Figure 5a). The restriction to two final states allows to determine whether particles containing both viral and cellular RNAs are favored as compared to particles containing only cellular RNAs, as input parameters or environmental conditions are modulated. This choice of a bimodal model is consistent with results of more complex calculations including polydispersity effects[23] . The titration curve of figure 5b shows that the concentration of particles containing large viral RNA is dominant as the concentration of large RNA is increased, illustrating the entropic selection mechanism mentioned above. Interestingly, it was found by Rulli and co-workers that some cellular RNAs detected in VLPs, in replacement of viral RNA, have indeed large length (about 7000 nt)[8]. This observation fits nicely the proposed mechanism of entropic RNA selection: incorporation of larger RNAs, such as messenger RNA, allows VLP assembly and the production of a larger amount of particles at constant cellular RNA amounts. In this case, the entropic selection mechanism will therefore strengthen the effect of the specific packaging signal $\psi$, that has been shown to be the main driving force of specific RNA uptake[8].

Similar calculations can be performed in the case where the two final states represent two particles with both different sizes and different RNA content. Specializing the previous model to small particles containing large viral RNA and cellular RNAs, and large particle containing mainly cellular RNAs, one finds again that small particles will dominate over large particles once a critical large RNA concentration has been reached (Figure 6). This behavior is qualitatively understood as follows. Considering first the case without RNA, if the enthalpy of formation of each capsid is only weakly selective in terms of capsid size, smaller capsids will be preferred for entropic reasons because they can be more numerous at fixed amount of proteins. The combination of this effect to the previous one of entropic selection of large genome leads to the prediction that small particles incorporating large RNA will be the dominant product of capsid self-assembly, and that the sizes of capsids are shifted to smaller values. This is nicely illustrated by comparing in figure 6b the shift of the boundary separating large RNA/no large RNA populations towards smaller large RNA volume fraction.

**Discussion**

Using Atomic Force Microscopy in order to image purified HIV-1 VLP produced by transfected human cells, we were able to quantify size distributions of VLP. The relevance of these size distributions was checked by performing the following control experiments. Indeed, the current purification scheme based on sucrose gradient is efficient in order to collect VLPs, but it has the side-effect of letting through small cellular vesicles of similar density called "exosomes"[26]. These are natural secretion products of cells under the present growth conditions. We quantified the size distribution of these additional cellular particles using transfection assays with a "mock" plasmid that is not coding for Gag or other viral proteins ("mock" conditions). Since the majority of these cell particles have sizes smaller than 90nm (Figure S1 in the supplemental material) and since VLP are not expected in this range, it is possible to get rid of the statistical contribution of exosomes within VLP purification conditions by imposing a lower size threshold of 90nm, as it is shown in figure S1 in supplementary material. Thus we are confident that in our study we mainly consider from a statistical point of view HIV-1 VLP and not exosomes. Moreover, the observed effect of RNA control of HIV-1 size polydispersity discussed below survives if exosome particles are kept in the analysis (cf Figure S2 in supplementary material).
We were also able to quantify the size distribution of VLP in the presence or in the absence of viral $\psi$-RNA. Our results show that these distributions are significantly different from a statistical point of view. Since the particles are produced in the same cell batch, the observed difference in size distribution is likely due to the direct influence of the viral RNA presence or absence, on the global self-assembly pathway. Up to our knowledge, this is the first time such an effect has been quantified on HIV-1 particles. This constitutes the main message of our work. Interestingly, a similar modulation of particle size and morphology has been recently evidenced within *in vitro* self-assembly on a plant RNA virus, the Cowpea Chloritic Mottle Virus (CCMV), by mixing purified capsid proteins with RNAs of controlled length[4,27] . In these works, it is shown that viral particles of CCMV are able to incorporate multiple copies of exogenous RNA which are smaller than the wild-type genome. Moreover, these particles are more heterogeneous in size than wild-type viral particles. This shows directly that the RNA content for this virus is able to modulate the size distribution of viral particles, similarly to our own observations with HIV-1. Additionally, recent observations on members of the paramyxoviruses family, such as the Newcastle Disease Virus also claim directly for a modulation of viral particle size by the stoichiometry of viral genome [28]. Indeed, it was observed in this case that a majority of infectious VLPs are small and contain a single genome, while a minority are large and contain multiple genomes.

In the case of HIV-1, the situation is necessarily more complex, due to the inherent greater complexity of the self-assembly scenario and to the lack of experimental control of molecular ingredients within cells during viral assembly, as compared to the framework of controlled *in vitro* self-assembly experiments by Cadena-Nava and co-workers[4] . However, it has been shown that RNA plays a role in retroviral assembly[7,29,30]. Moreover, retroviruses incorporate viral RNA as well as cellular RNA of different sizes and categories (mRNA, tRNA and 7S RNA for example). In the absence of viral genomic RNA, retroviruses compensate the lack mainly by incorporating cellular mRNA[7,8]. But none of these studies have shown a difference of morphogenesis in retroviral assembly of VLP with or without viral RNA. Our work suggests that the difference lies essentially in the size distribution of VLP, which is significantly smaller in the presence of viral RNA, and more homogeneous.

     Using simple argument of micellization thermodynamics, we were able to reproduce qualitatively the previous observations of size modulation by the presence of viral RNAs. In particular, for viral particles with equal sizes but different RNA content, there is an entropic selection of large viral RNA. Moreover, for viral particles with both unequal sizes and different RNA content, there is an entropic selection of both small particle size and large viral RNA. In order to obtain these results, a number of simplifying assumptions have been obviously made, and therefore are likely to prohibit a quantitative comparison between the model and the data. One of the strongest assumption is that the energetic cost of particle formation is only loosely size selective, allowing therefore entropic effects to dominate in the model. In the case of HIV-1, this assumption is realistic, since we measured that VLPs have indeed a large size dispersion. The reason for such a large dispersion is not known, but it might be related to the recently observed incompleteness of the protein layer underlying the membrane envelope: indeed the weak size dispersion of non-enveloped viruses has been recently linked to the bending cost of the complete and ordered protein layer within the capsid, while for incomplete protein layers of non-envelope this cost should be strongly reduced and therefore the control of size dispersion through bending cost should not be efficient[19]. Interestingly, self-assembly of Gag proteins on monodisperse gold particles covered by nucleic acids has been recently shown to produce VLPs with strongly reduced polydispersity, showing therefore that the size dispersion is mainly determined by the nucleic acid content of VLPs[31].
     One other potential pitfall of the model is that HIV-1 self-assembly is a rich and complicated process for which it is not clear whether equilibrium considerations like the one underlying our model can apply. However, even within the framework of out-of-equilibrium self-assembly, the entropic arguments should still hold as the number of configurations of the system is drastically changed upon self-assembly. Despite these limitations, it is remarkable that such a simple model is able to qualitatively explain the role of viral RNA in HIV-1 assembly at regulating the size distribution of viral particles.

**Material and methods**

*Cell culture*
Human 293T cells used in this study were maintained in Dulbecco's modified essential medium (DMEM) supplemented with 10% fetal calf serum (FCS) and antibiotics (penicillin/streptomycin).

*DNA Plasmids*
The plasmid expressing HIV-1 Gag alone (pGag) and the plasmid expressing HIV-1 Gag, GagPol (lentiviral vector named pCMVΔ8.91) were respectively described in Burniston *et al.*[32] and Zufferey *et al.*[21]. Another plasmid is expressing a non infectious HIV-1 « ψ » mRNA, as described by Naldini *et al*[22]. An empty vector (pcDNA3.1 expressing the fluorescent protein Venus) is used as a control for "Mock" transfected cells.

*Antibodies*
Antibodies used in this study are a mouse anti-CAp24 and a rabbit anti-MAp17 [NIH, USA], the rabbit anti-IN was a gift of Dr JF.Mouscadet (LBPA, ENS Cachan), the rabbit anti-RT was a gift of Dr JL.Darlix (ENS Lyon). The anti-mouse and anti-rabbit antibodies coupled to horseradish peroxidase [Dako ] were used for immunoblot detection.

*Particles production and purification*
To obtain non-infectious HIV-1 VLPs, human HEK 293T cells were transfected with the plasmid p8.91 that encodes the polyproteins Gag and Gag-Pol, as well as for the regulatory viral proteins Tat and Rev. The resulting VLPs give rise to mature viral particles in which Gag is cleaved by the viral protease and the core is formed. Immature VLPs were produced by a plasmid encoding only Gag and thus yielding viral particles in which the Gag remains uncleaved. All samples were produced in the presence or in the absence of a plasmid expressing the HIV-1 ψRNA[22]. The "mock" condition is obtained by cell transfection with a "mock" plasmid that is not coding for viral Gag or any other viral proteins.
The DNA plasmid-phosphate transfection complex was produced by mixing a phosphate buffer (HBS) with the corresponding plasmids previously supplemented with CaCl2 (250 mM), and then added on plated cells. Cells were incubated at 37 °C overnight. Then, cells were washed with PBS and new medium was added for 24 hours of virus production. The medium containing viruses was collected and clarified by filtration (0,45 μm). Then, it was deposit at the top of a double layer sucrose gradient of 20% and 30% sucrose in PBS and submitted to an ultracentrifugation in a Beckman SW41 rotor at 27000 rpm for 2 hours at 4°C. VLPs were isolated at the bottom. Viral cores were isolated by adding the detergent Triton X-100 (0.05%) to the upper sucrose layer. Viral particles or cores were diluted in TNE (10 mM Tris pH 7.4, 100 mM NaCl, 1 mM EDTA, pH 7.4), at 4°C for 2 hours and directly deposited on an AFM surface. This technique is adapted from Accola *et al*[33].

*Immunoblotting*
To analyze viral proteins by immunoblot analysis, the proteins contained **in** the viral pellets (VLP or cores) were separated onto an SDS–10% PAGE gel and transferred onto a polyvinylidene difluoride membrane (Hybond-P, Amersham), and immunoblotting was performed by using the corresponding antibodies, as described previously[34]. Secondary antibodies were revealed by using the SuperSignal West Pico substrate (Thermo Scientific). The signals were revealed by autoradiography on Fuji film.

*Reverse transcription assay*
Ten microliters of supernatant containing virus-like particles was added to 50 μl of a reverse transcription mix (60 mM Tris pH 8.0, 180 mM KCl, 6 mM $MgCl_2$, 0.6 mM EGTA pH8.0, 0.12% Triton X-100, 6 mM dithiothreitol, 6 µg/mL oligo dT, 12 µg/ml poly rA, and 20 µCi/ml [α-32P]dTTP [specific activity, 3,000 Ci/mmol]). After 1 h of incubation at 37°C, 5 µl was loaded onto DEAE paper (DE-81; Whatman) and then rinsed with 2× SSC (0.3 M NaCl, 0.03 M sodium citrate [pH 5]). The radioactivity (X-rays) of the sample was recorded by a storage phosphorscreen (Molecular Dynamics), measured with a phosphorimager (Fuji), and quantified by using MultiGauge software (Fuji).

*Atomic Force Microscopy imaging*
An atomically flat mica (grade V-I, SPI) surface functionalized with positively charged poly-L-lysine (Sigma Aldrich) was chosen to assure the attachment of viral particles for AFM imaging: 10 µl poly- L-lysine (0,01%) were deposited onto freshly cleaved mica-discs. After an incubation of 30 seconds, disks were rinsed twice with 1 ml of ultrapure water and dried by a nitrogen flow. A 5 µl-droplet of the viral agent obtained after purification was applied to the surface and incubated for two minutes (VLPs) or 15 minutes (viral cores) to assure a sufficient adsorption.
Imaging was performed in Tapping Mode with a Nanoscope IIIa Multimode AFM operated with a JVH- scanner (Digital Instruments Veeco, Santa Barbara). Since our main objective in this work is to compare size distributions of VLPs under different production conditions, we chose to image dried samples, although this might flatten particles, in order to avoid desorption of particles during scanning of the surface by the AFM tip. For imaging in air, the sample was rinsed with 1 ml of ultrapure water and gently dried by a nitrogen flow. Imaging was executed with silicon tips (resonance frequency ~ 350 kHz, $k_{cantilever}$ = 40 N/m). According to the scan size of 0.5-3 µm, a scan rate between 1 and 3 Hz was chosen. Additional AFM imaging in liquid conditions are provided as supplemental data.

*Automated image analysis*
Images acquired by AFM were analyzed in an automated fashion using home-made MATLAB (from Mathworks) scripts. The principle of this analysis is summarized in the figure 2. The aim of developing such automated image analysis is to be able to select particles based on objective criteria. While using this approach, caution should be exercised in order to optimize the parameters of the filters, allowing to reject aberrant particles (broken particles, aggregates...) and select normal particles, while keeping size dispersion and good statistics.
The first step of the image analysis is to use standard flattening procedure of the raw AFM image in order to get rid of typical piezo-electric drift during image acquisition. The next step is to use a low height selection (12nm in our case) in order to focus on image parts above the roughness of the naked substrate (mica+Poly-L-Lysine coating) and to make a binary image out of it. Then selection is made on image parts that enter a certain area range. For VLPs, an optimal choice for this range is between 7000nm$^2$ (equivalent to 94nm diameter) and 10$^5$ nm$^2$ (equivalent to 357nm diameter). Another height selection is used to ensure that the selected parts of the image reach a minimal height (here 25nm) compatible with the *a priori* size of the particle to be focused on. Then particles partly lying on one edge of the image are removed. Finally, the actual shape of the binary particles on the image is used to extract the perimeter $P_i$, the area $A_i$, the major axis $M_i$ and the minor axis $m_i$. This information is used in order to perform a last selection step with one morphological parameter: the fractal parameter $F_i = \frac{4\pi A_i}{P_i^2}$. The fractal parameter allows to focusing on particles that have smoothed contour (0.75<$F_i$<1) or shaky contour ($F_i$<0.75). In particular, this parameter is easily optimized in order to reject multiple overlapping particles that went through the area selection step. For an ideal disk, the value of fractal parameter is 1. The final binary image is then multiplied by the original image and is now ready for the computation of statistical properties of the particles.

*Model of self-assembly process*

The self-assembly of particles and the influence of the RNA content is modelized thanks to a classical description of micellization[18,19,24,27,35]. The aim of the model is to predict the partitioning of different molecules (identical proteins, identical large RNAs, and identical small RNAs) into the products of self-assembly under well defined conditions and at equilibrium. In particular, we will seek for which conditions it is possible to reproduce qualitatively the experimental results. For the sake of simplicity, we present a bimodal version of the calculation, for which there are only two possible products (labelled 1 and 2). We recently published an extended analysis of the present model including the effect of particle polydispersity[23]. It turns out that the results of the present simplified bimodal analysis do reproduce quantitatively the one of the more complex model. This justifies our use of simplified bimodal model, as a pedagogical route in order to interpret the experimental results. The numerical calculations presented in the main text are based on the resolution of the equilibrium equations associated to the self-assembly. More information on this model are available in the supplemental data.

Assuming that initial concentrations of respectively proteins, large and small RNAs are noted $\phi_0, \phi_{r+}, \phi_{r-}$, the equations for the case of bimodal self-assembly are written as a combination of mass conservation laws and mass actions laws, as it is demonstrated in the supplemental data:

$$\phi_0 = c_0 + p_1 c_0^{p_1} c_{r+}^{n_1} c_{r-}^{m_1} e^{-p_1 g_1} v_0^{p_1+n_1+m_1-1} + p_2 c_0^{p_2} c_{r+}^{n_2} c_{r-}^{m_2} e^{-p_2 g_2} v_0^{p_2+n_2+m_2-1}$$

$$\phi_{r+} = c_{r+} + n_1 c_0^{p_1} c_{r+}^{n_1} c_{r-}^{m_1} e^{-p_1 g_1} v_0^{p_1+n_1+m_1-1} + n_2 c_0^{p_2} c_{r+}^{n_2} c_{r-}^{m_2} e^{-p_2 g_2} v_0^{p_2+n_2+m_2-1}$$

$$\phi_{r-} = c_{r-} + m_1 c_0^{p_1} c_{r+}^{n_1} c_{r-}^{m_1} e^{-p_1 g_1} v_0^{p_1+n_1+m_1-1} + m_2 c_0^{p_2} c_{r+}^{n_2} c_{r-}^{m_2} e^{-p_2 g_2} v_0^{p_2+n_2+m_2-1}$$

In these equations, $c_0, c_{r+}, c_{r-}$ are respectively the concentration of proteins, large RNA and small RNA that remain in solution as monomeric species. The particles 1 and 2 have respectively $p_1$ and $p_2$ proteins, $n_1$ and $n_2$ large RNAs, and $m_1$ and $m_2$ small RNAs. The free energy of formation of particle 1 and 2 per protein are respectively $g_1$ and $g_2$ in units of thermal energy $k_B T$. Finally the quantity $v_0$ is a typical volume allowing to use dimensionless concentrations in the equations. An additional constraint is imposed on the relation between number of proteins and RNAs, as it is discussed in the supplementary material. The two relations are written as

$$p_1 = K_v n_1 + K_c m_1$$

$$p_2 = K_v n_2 + K_c m_2$$

, where $K_v$ and $K_c$ are respectively proportional to the length of viral and cellular RNAs. In order to show the general trends of the model, we used small numbers of proteins per aggregate for the sake of convenience. As a consequence, only relative concentrations are relevant to the discussion, and the absolute values of concentrations are not to be compared to real experimental situations. We checked that the results presented here are not strongly dependent on the precise values of parameters. The first parameters used to compute figure 5 are the following: $p_1 = 21, n_1 = 2, m_1 = 10, g_1 = -2$ ; $p_2 = 21, n_2 = 0, m_2 = 14, g_2 = -2$ ; $\phi_0 = 0.4, v_0 = 2, K_v = 3, K_c = 1.5$. The second set of parameters used to compute figure 6 are the following: $p_1 = 21, n_1 = 2, m_1 = 10, g_1 = -2$ ; $p_2 = 42, n_2 = 0, m_2 = 28, g_2 = -2$ ; $\phi_0 = 0.4, v_0 = 2, K_v = 3, K_c = 1.5$.

**Acknowledgements**

The authors would like to thank the *Fondation Simone et Cino Del Duca* from the *Institut de France* who allowed to launch this interdisciplinary project.

Supplementary Data are available online: supplementary figures 1-5 and supplementary methods.

Figure Captions

Figure 1: Biochemical characterization of VLPs and viral cores, and AFM imaging. *(a)* Immunoblot of HIV-1 for mature and immature VLPs and cores. *(b)* Reverse transcription test on mature VLPs and cores in the presence or the absence of ψRNA in the VLPs or cores. *(c)* Typical images of mature VLPs. *(d)* Typical images of viral cores.

Figure 2: Automated image analysis. (a) Example of the successive image analysis steps applied on a single image. (b) Cartoon of the principle of image analysis. (c) Various representation (color map, contour plots, 3D plots) of particles that were selected. *Top* VLP, *down* core. Interestingly, the contour map of the VLP shows a pronounced asymmetry close to its maximal height, reflecting the presence of an asymmetric object inside the VLP.

Figure 3: Statistical analysis of size distributions of VLPs and cores obtained through automated image analysis. The number of particles is indicated by the value $N$. *(a)* and *(c)* 2D histograms of short and long diameters for respectively mature VLPs and cores. *(b)* Short diameter histogram obtained by projecting the 2D histograms for VLPs and cores. *(d)* and *(f)*, long diameter histogram obtained projecting the 2D histograms for VLPs and cores. *(e)* Typical example of short and long diameters measured on a single particle.

Figure 4: Influence of the presence or absence of viral ψRNA on particle morphogenesis. The number of particles is indicated by the value $N$. *(a)* and *(c)* 2D histogram of short and long diameters for respectively VLPs in the absence and in the presence of ψRNA. The size distribution is shifted toward smaller value, and its dispersion is reduced. *(b)* and *(d)* 2D histogram of short and long diameters for respectively cores in the absence and in the presence of ψRNA. The same shift in the distribution is observed, although with w weaker amplitude. *(e)* Box plot of equivalent diameters summarizing the previous results. The equivalent diameter is obtained by converting the 2D projected area of the particle into the diameter of a disk that would give the same area.

Figure 5: Model of entropic selection of viral genome at fixed particle size. *(a)* Cartoon of the self-assembly. The model is specialized to bimodal products of self-assembly, in which the size of particle are equal and the RNA content are different. (b) Typical large RNA titration computed thanks to the model detailed in the *supplemental data*. The value of parameters chosen for the calculation are found in the *material and methods*. Blue circles correspond to particles with large RNA, and green crosses correspond to particles lacking large RNA.(c) Phase diagram *small RNA / large RNA*. The boundary for which the concentration of both particles are equal is shown by blue filled circles. The line joining the circles is drawn to guide the eye.

Figure 6: Combined model of viral genome and particle size entropic selection. The model is specialized to bimodal products of self-assembly, with different particle size and different RNA content. *(a)* Typical large RNA titration computed thanks to the model detailed in the *supplemental data*. The value of parameters chosen for the calculation are found in the *material and methods*. In this case, the number of proteins in the "large particles" is twice the number of proteins in the small one. Red circles correspond to small particles with large RNA, and pink crosses correspond to large particles lacking large RNA. *(b)* Phase diagram *small RNA / large RNA*. The boundary for which the concentration of both particles are equal is shown by red filled circles. The line joining the circles is drawn to guide the eye. For comparison, the boundary at fixed particle sizes found in figure 5b is depicted by a blue dotted line.

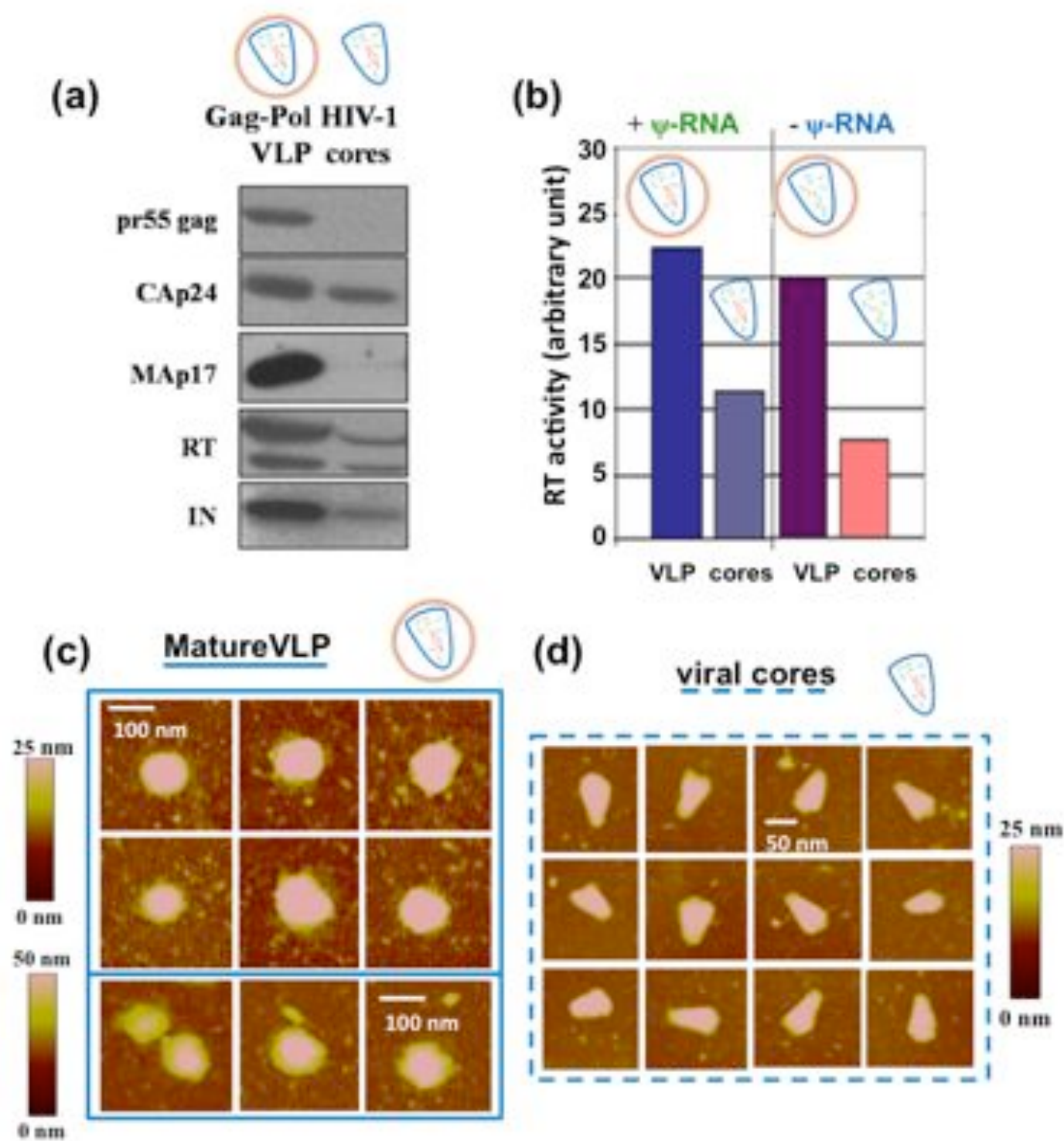

Figure 1

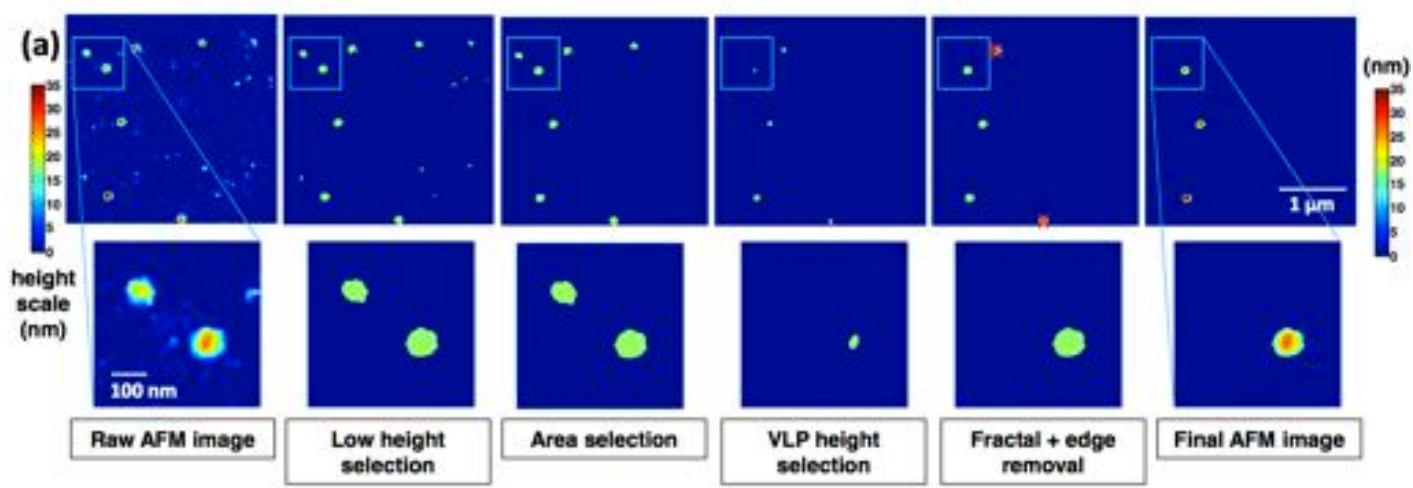
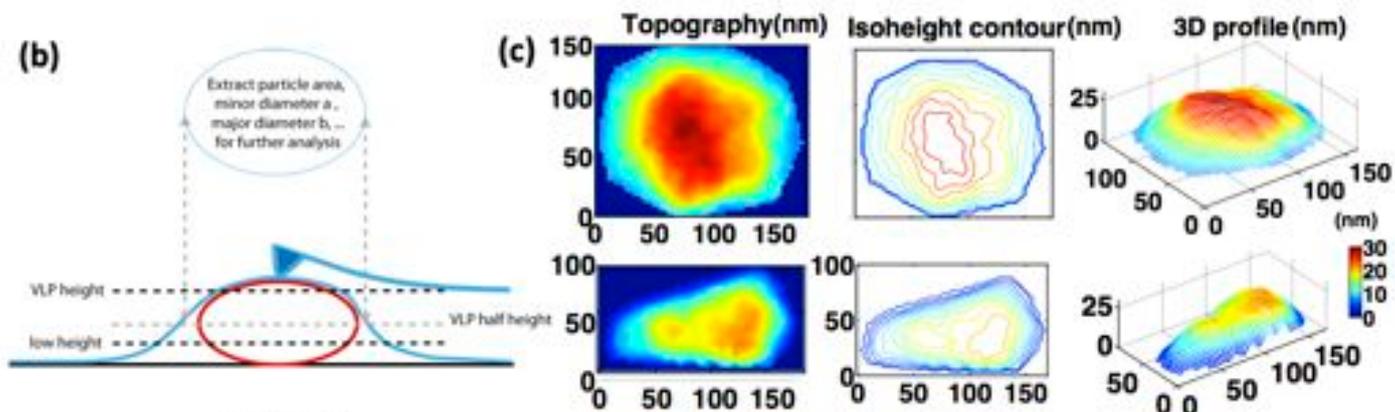

Figure 2

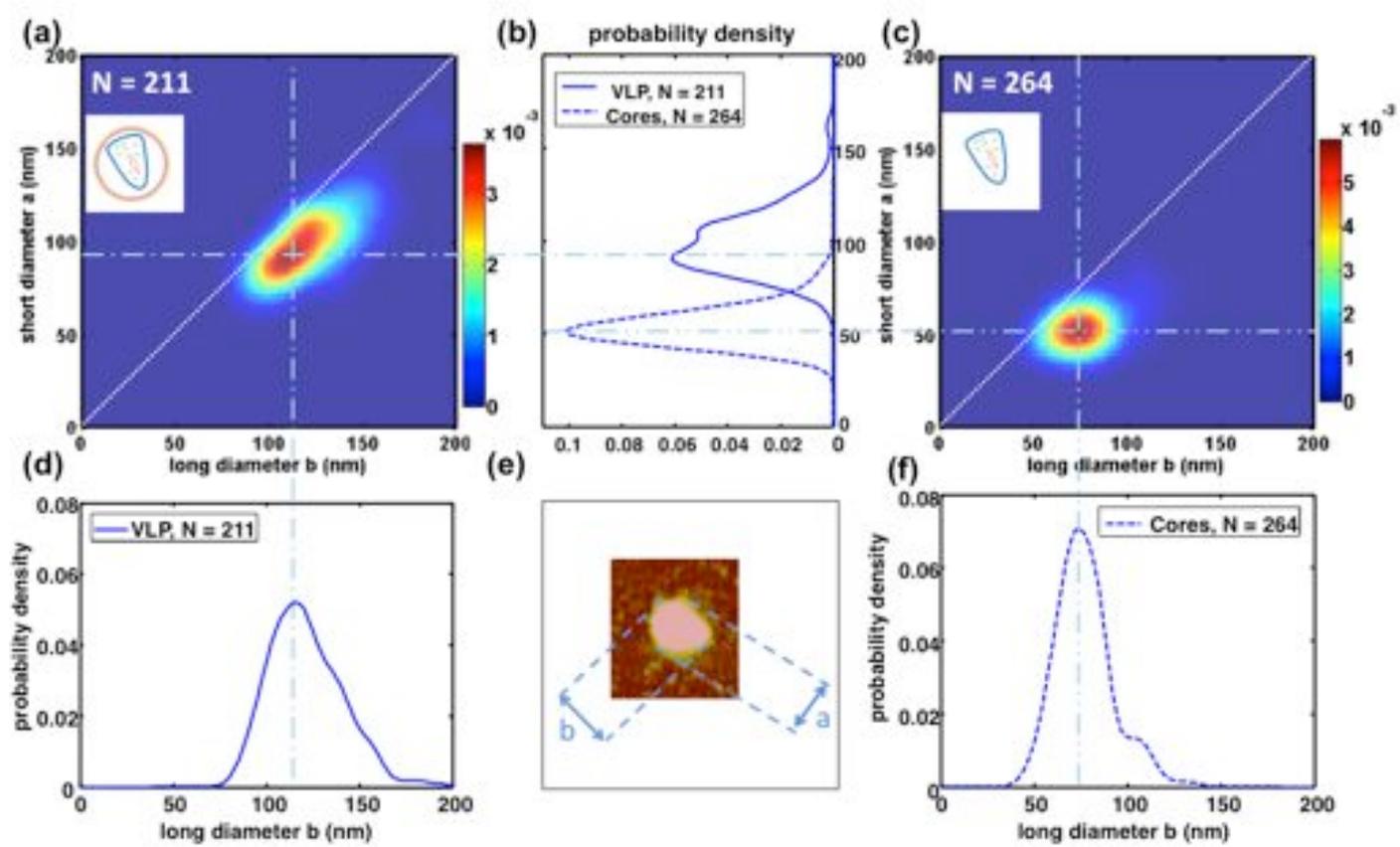

Figure 3

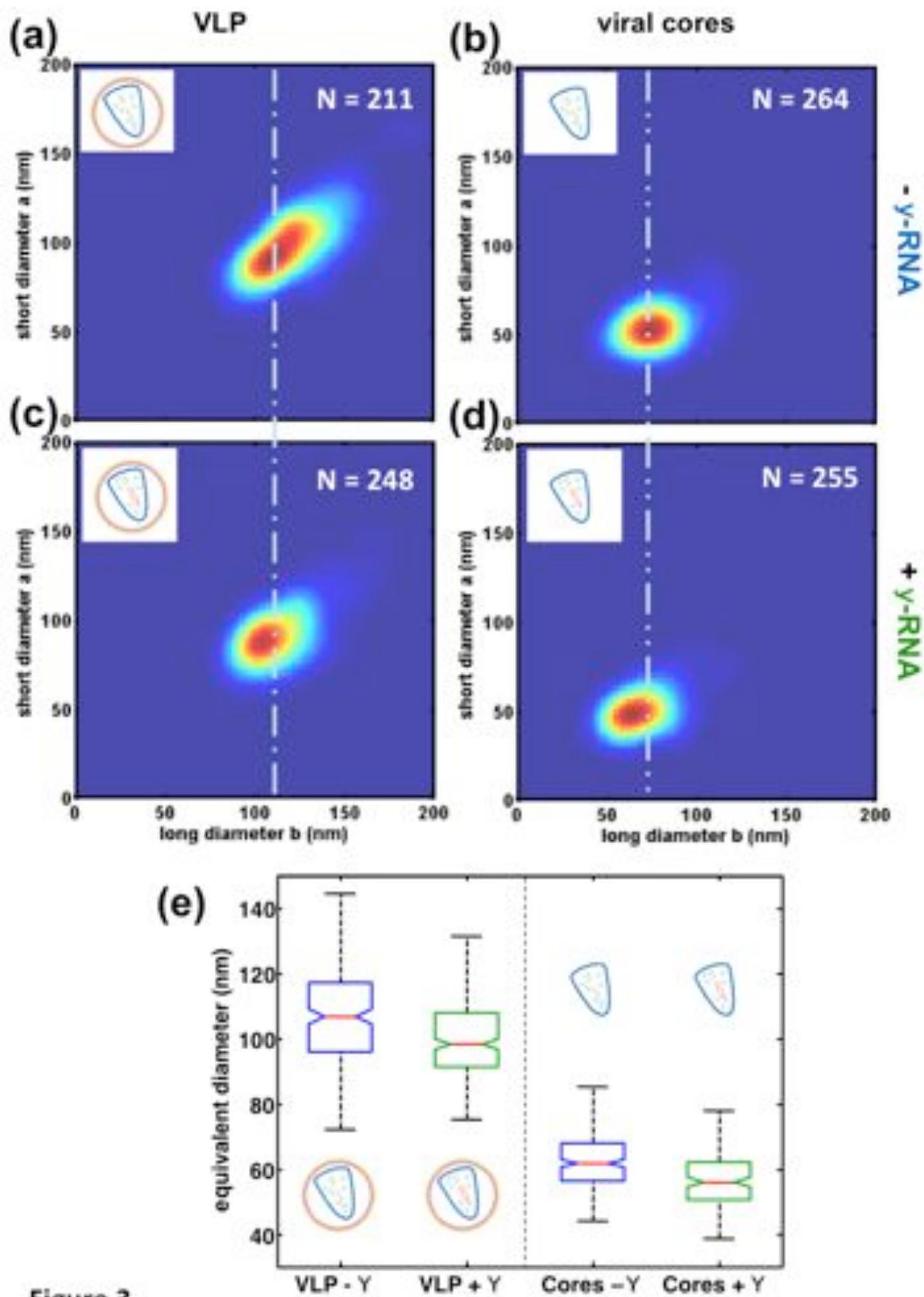

Figure 3

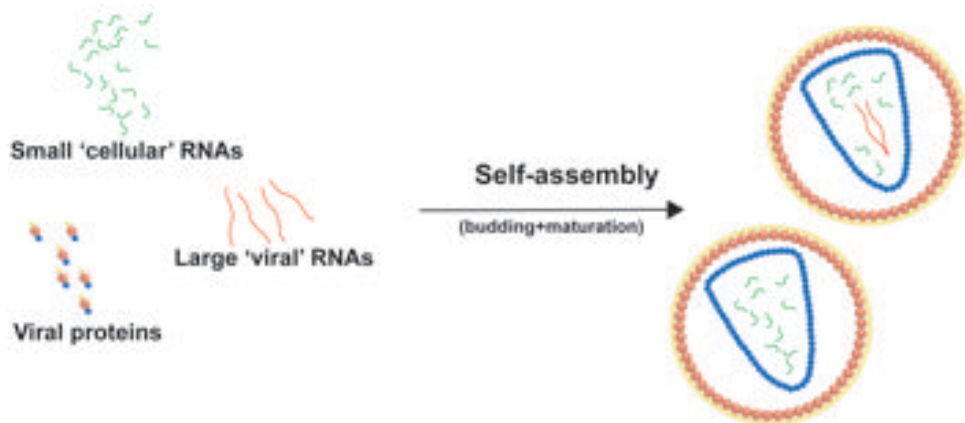
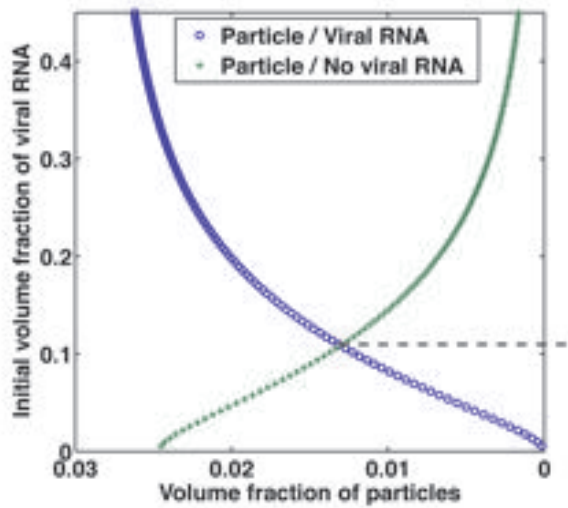
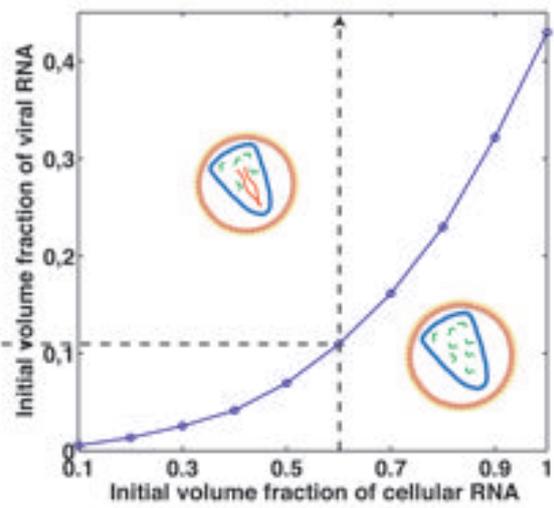

**Figure 5**

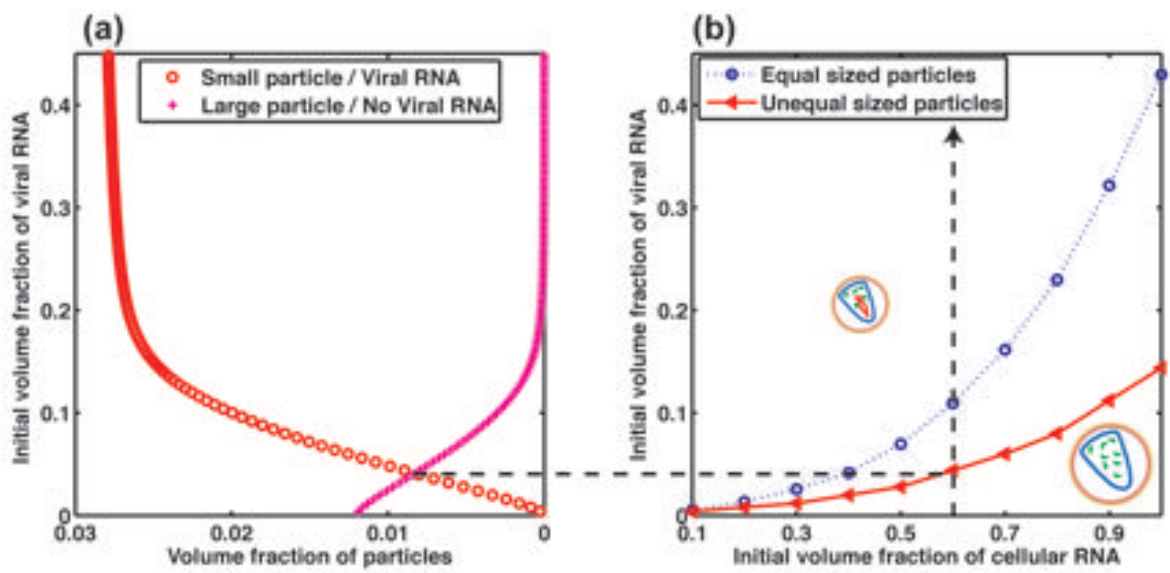

**Figure 6**